\documentclass[aps,prb,supersrciptaddress,twocolumn,showpacs,amsmath,floatfix,citeautoscript]{revtex4-2}

\usepackage{amsmath}
\usepackage{graphicx}
\usepackage{epstopdf}
\usepackage{subfigure}
\usepackage{braket}
\usepackage{color}
\usepackage{epstopdf}
\usepackage{gensymb}
\usepackage{amssymb}
\usepackage{float}

\begin{document}
\title{First-principles study of magnetic structures of triangular antiferromagnets NaYbS$_2$ and NaYbO$_2$}

\author{Da-Ye Zheng}
\affiliation{CAS Key Laboratory of Quantum Information, University of Science and Technology of China, Hefei 230026, Anhui, China}
\affiliation{Synergetic Innovation Center of Quantum Information and Quantum Physics, University of Science and Technology of China, Hefei, 230026, China}
\author{Zhen-Xiong Shen}
\affiliation{CAS Key Laboratory of Quantum Information, University of Science and Technology of China, Hefei 230026, Anhui, China}
\affiliation{Synergetic Innovation Center of Quantum Information and Quantum Physics, University of Science and Technology of China, Hefei, 230026, China}
\author{Meng Zhang}
\affiliation{CAS Key Laboratory of Quantum Information, University of Science and Technology of China, Hefei 230026, Anhui, China}
\affiliation{Synergetic Innovation Center of Quantum Information and Quantum Physics, University of Science and Technology of China, Hefei, 230026, China}
%\author{Xinguo Ren}
%\affiliation{xxxxxxxxxxxxxxxxxxxxxxxxxxxxx}
\author{Lixin He}
\email{helx@ustc.edu.cn}
\affiliation{CAS Key Laboratory of Quantum Information, University of Science and Technology of China, Hefei 230026, Anhui, China}
\affiliation{Synergetic Innovation Center of Quantum Information and Quantum Physics, University of Science and Technology of China, Hefei, 230026, China}

\date{\today }

\begin{abstract}
We investigate the magnetic interactions in triangular rare-earth delafossites materials NaYbO$_2$ and NaYbS$_2$ via
first-principles calculations. The calculated Curie-Weiss temperatures are in good agreement with experiments.
We perform classical Monte Carlo simulations of the two compounds using the extracted
exchange parameters. We find that if only the nearest neighbor interactions are considered, the magnetic ground states of NaYbO$_2$ and NaYbS$_2$ are a stripe and a planar 120\degree~ N\'{e}el state, respectively.
The simulated transition temperatures are much higher than the lowest experimental temperatures, where no magnetic ordering was observed.
However, we show by adding suitable second neighbor interactions, the {\it classical} magnetic ground state of NaYbO$_2$ becomes to the $Z_2$ vortex phase,
and the simulated specific heat $C_v$ are very similar to the experimental observations,
with no obvious phase transition down to the extremely low temperature.

\end{abstract}

\maketitle

\section{INTRODUCTION}

Quantum spin liquids (QSL) are exotic states of matter,
in which strong frustration and the quantum fluctuations
prevent long-range magnetic ordering down to the
zero temperature~\cite{Anderson1973,Anderson1987,Balents2010,Savary2017}.
The QSL states are highly entangled, with novel excited state properties,
such as emergent gauge fields and fractional particle excitations\cite{Balents2010,Savary2017},
and therefore has attracted great attention since the concept had
been proposed by Anderson~\cite{Anderson1973}.

One of the promising routines to search for the QSL is in
the geometrically frustrated materials\cite{frustrated_magnetism_book}.
Recently, rare earth triangular lattice materials YbMgGaO$_4$ has been proposed to host a gapless QSL with effective-spin-$1/2$ local moments,
in which no sign of long-range spin ordering has been observed down to the lowest temperature, $T$$\approx$50 mK by various techniques~\cite{liyuesheng2015-prl,liyaodong2016,Li2016,Shen2016,Paddison2017}.
However, it has been argued that the disordered spin state in YbMgGaO$_4$ might come from the Ga/Mg disorder, instead of true QSL~\cite{Paddison2017,Zhu2017,Parker2018}.
On the other hand, the rare-earth delafossites AReCh$_2$ materials have perfect triangular layers, without the lattice distortion and site mixing in YbMgGaO$_4$.
%, and therefore ideal for studying the QSL.
Very recently, high quality samples of a large family of AReCh$_2$ materials ~\cite{liu2018,xing2019,Ranjith2019}, including NaYbS$_2$ \cite{Sarkar2019} and NaYbO$_2$ \cite{Bordelon2019} have been synthesized.
%It has been demonstrated that the NaYbCh$_2$ compounds have
No magnetic ordering or transition in these materials are observed down to extremely low temperature
from specific heat and susceptibility measurement.
Therefore, they are ideal candidates for searching QSL.

It is well known that the ground state of the isotropic Heisenberg model on a triangular lattice is a planar 120\degree~ N\'{e}el state, instead of QSL~\cite{Capriotti1999}.
%To explain the possible QSL behavior in the triangular lattice,
Li et. al. proposed that the spin-orbit interactions
may introduce anisotropic exchange interactions, which may add frustration to the model, leading to the QSL ground state \cite{liyuesheng2015-prl,liyaodong2016}.

In this work, we investigate the magnetic properties of NaYbO$_2$ and NaYbS$_2$.
We calculate the magnetic interactions in these materials, via first-principles calculations,
and fit them to the anisotropic exchange model.
The calculated Curie-Weiss temperatures are in good agreement with experiments.
We then perform classical Monte Carlo simulations of the two materials, using the obtained exchange parameters, to investigate their finite temperature behaviors. We find that the ground state of NaYbS$_2$ is the planar 120\degree~ N\'{e}el state, whereas the ground state of NaYbO$_2$ has a stripe order, if only the nearest neighbor interactions are considered.
The simulated transition temperatures are much higher than the lowest experimental temperatures, where no magnetic ordering was observed.
However, we find that by adding suitable second nearest neighboring exchange interactions, the magnetic ground state becomes the $Z_2$ vortex.
The simulated specific heat $C_v$ are very similar to the experimental observations,
with no obvious phase transition down to the extremely low temperature.

\section{Methods}

The electronic and magnetic properties are calculated via density functional theory,
within the generalized gradient approximation of the Perdew-Burke-Ernzerhof (PBE)\cite{pbe-ref},
implemented in Vienna ab initio simulations package (VASP)~\cite{vasp-ref}.
The projector-augmented wave (PAW) pseudopotentials with spin-orbit couplings (SOC) are used.
%where the Yb 4$f$6$s$, Na 3$s$ , O 2$p$, S 3$p$
%electrons are treated as valence electrons.
A 500 eV plane-wave energy cutoff results in very food convergence.
The on-site Coulomb interactions $U$-$J$=6.0 eV are included for Yb 4$f$ electrons in a rotationally invariant
scheme~\cite{Dudarev1998}.
%\blue{We also do the calculations using the rotationally invariant DFT+U method by Liechtenstein et al.\cite{Liechtenstein1995}, and obtain very similar results for different $U$ and $J$ parameter values.}
The experimental crystal structures are used for the calculation \cite{NaYbS2-stru-ref, NaYbO2-stru-ref}.
For calculations using primitive unit cells, an 11$\times$11$\times$11 $k$-point mesh is used,
whereas a 9$\times$9$\times$4 $k$-point mesh is used for the calculations on the conventional unit cell,
containing 48 atoms.

\section{Results and Discussion}

\subsection{Crystal structures}

The NaReCh$_2$ (where Ch=O, S and Re=Yb, Ga, Tb are the rare-earth ions) is a large family of materials,
which have an ideal triangular lattice structure, with space group $R\bar{3}m$.
In this work, we focus on the properties of two representative compounds: NaYbS$_2$, and NaYbO$_2$.

\begin{figure}[tp]
\centering
\includegraphics[width=0.45\textwidth]{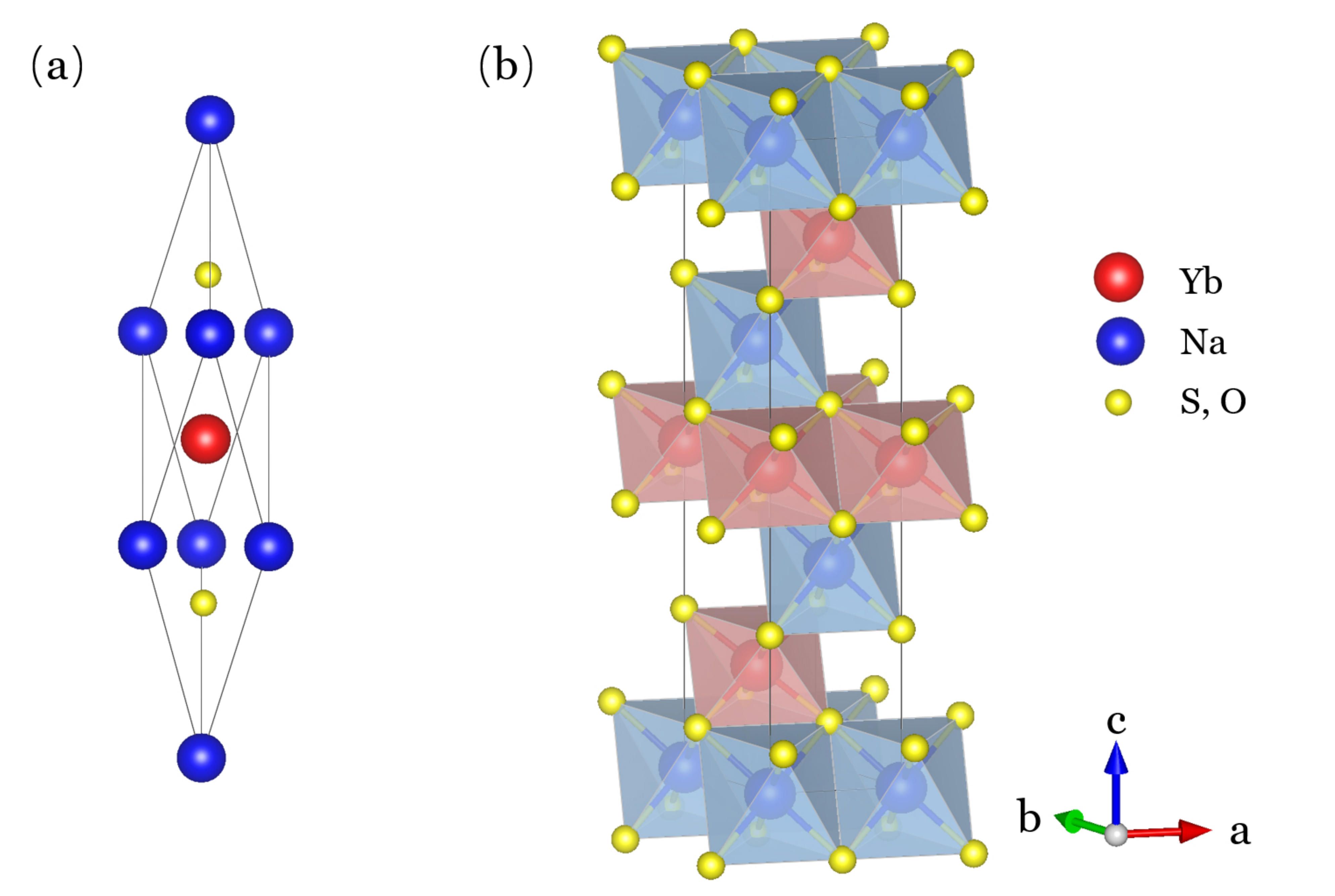}
\caption{(Color online) (a) The primitive unite cell and (b) a
 $\sqrt{3}$$\times$$\sqrt{3}$$\times$1 conventional unit cell of NaYbS$_2$.
The Yb$^{3+}$ ions are located in the center of YbS$_6$ octahedrons, which have an ABAB
stacking along the $c$ axis, separated by the Na layers.}
\label{fig:structure}
\end{figure}

Figure~\ref{fig:structure}(a) depicts the structure of the primitive unit cell of NaYbS$_2$,
containing four atoms, whereas a $\sqrt{3}$$\times$$\sqrt{3}$$\times$1 conventional unit cell
%(where the $c$-axis is along the [110] direction of the primitive unit cell),
is shown in Fig.~\ref{fig:structure}(b).
The Yb$^{3+}$ ion and its six surrounding S ions form a YbS$_6$ octahedron.
The Yb$^{3+}$ ions are located at the centers of the octahedrons, which are
the centers of the D$_{3d}$ symmetry~\cite{Bordelon2019}, precluding any Dzyaloshinskii–Moriya \cite{dzyaloshinskii64,moriya60}
distortions. Indeed, the Na NMR lines reveal an absence of inherent structural distortions
in NaYbS$_2$~\cite{Baenitz2018}, and NaYbO$_2$~\cite{Bordelon2019}.
Therefore, the Yb$^{3+}$ ions form a perfect triangular quasi-2D lattice. This is in strong contrast to the well-explored spin-liquid candidate YbMgGaO$_4$~\cite{liyuesheng2015-prl,liyaodong2016,Li2016,Shen2016,Paddison2017},
which has considerable site mixing of the Ga and Mg ions.
The YbS$_6$ octahedrons have an ABAB
stacking along the $c$ axis, separated by the Na layers,
and the magnetic coupling
between different Yb$^{3+}$ layers are expected to be negligible.
The lattice constants of NaYbS$_2$ are $a$=3.901 \AA ~and $c$=19.736 \AA~\cite{NaYbS2-stru-ref}, which are much larger than those of NaYbO$_2$, $a$=3.346 \AA  ~and $c$=16.456 \AA~\cite{NaYbO2-stru-ref}.

\subsection{Band structures}

\begin{figure}[htb]
\centering
\includegraphics[width=0.48\textwidth]{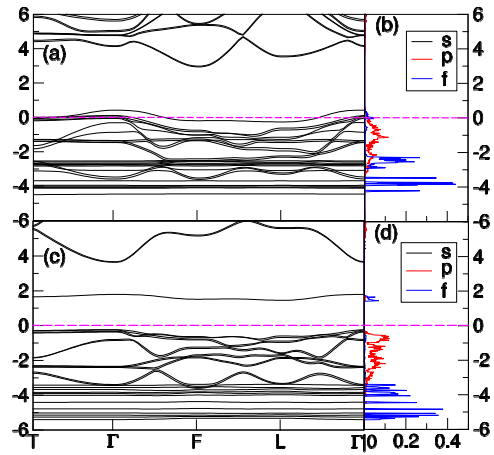}
\caption{(Color online) Left: The band structures of (a) NaYbS$_2$ and (c) NaYbO$_2$.
Right: The corresponding PDOS of Na 3$s$, S 3$p$ (O 2$p$), and Yb 4$f$ orbitals
are shown in (b) and (c) respectively. The dashed line denotes the Fermi level.
}
\label{fig:bandstructures}
\end{figure}

Figure~\ref{fig:bandstructures}(a),(c) depict the electrical bands structures of NaYbS$_2$ and NaYbO$_2$ respectively, and the corresponding partial density of states (PDOS) for Na 3$s$, Yb 4$f$
and S/O $p$ electrons are shown in Fig.~\ref{fig:bandstructures}(b),(d). The PDOS of Na 3$s$ electrons
are marginal in this energy window as shown in the figures.
The band structures and PDOS are calculated using  primitive unit cells with the FM spin configuration,
i.e., all spin of Yb$^{3+}$ ions are forced to align along the $z$-axis, with SOC turned on.
The Fermi levels are dominated by the S/O $p$ orbitals. The Yb 4$f$ electrons form
rather flat bands, which are about 2 - 4 eV below the Fermi level for NaYbS$_2$,
and about 4 - 6 eV below the Fermi level for NaYbO$_2$. These results suggest that the Yb 4$f$ states are
very localized, with only small hybridization between the Yb 4$f$ and S/O $p$ electrons.
But still, one can see that the Yb 4$f$ electrons hybridize stronger with the S $p$ orbitals than with the
O $p$ orbitals.

Experimentally, NaYbS$_2$ and NaYbO$_2$ are insulators, with band gaps equal 2.7 eV and 4.5 eV respectively~\cite{liu2018}.
However, there are about 0.83 electrons per unit cell above the Fermi level for NaYbS$_2$ calculated by the DFT+U method. The NaYbO$_2$ turns out to be an insulator from the DFT+U calculations,
however, the calculated bandgap is 1.70 eV, which is also significantly smaller than the experimental values.
These results suggest that NaYbS$_2$ and NaYbO$_2$ are strongly correlated materials, and may not be well described
by the DFT+U method, which treats the on-site Coulomb $U$ in an over-simplified mean-field way.
To accurately calculate the electronic structure of  NaYbS$_2$, and NaYbO$_2$  is an interesting and challenging problem, which may require more sophisticated many-particle techniques (e.g. dynamics mean-field theory\cite{DMFT1996}) to treat the strong correlation effects.
Despite this, we would still like to calculate the magnetic interactions in these materials to shed some light on possible QSL states in these materials.

\subsection{Magnetic exchange interactions}
\label{sec:exchange_parameters}

In NaYbS$_2$ and NaYbO$_2$, the 4$f$ electrons of Yb$^{3+}$ ions couple strongly to
the orbital momentum, resulting in a
total angular momentum $J$=7/2 state, which splits under the crystal field.
It has been shown that the ground state spin doublet is well separated from the excited spin doublets~\cite{liyaodong2016}, and therefore, the system can be treated as an effectively spin-1/2 system.
The strong SOC coupling in these materials further introduces anisotropic magnetic exchange interactions~\cite{Witczak2014,liyuesheng2015-prl,liyaodong2016}. We would first like to extract the magnetic exchange interactions as input parameters for further studies.

Li et al. derived a general Heisenberg model Hamiltonian based on symmetry consideration for the triangular compounds~\cite{liyaodong2016}.
The Hamiltonian reads, following the notation of Ref.\onlinecite{Maksimov2019},
\begin{align}
\begin{split}
\mathcal{H}=& \sum_{\langle i j\rangle}\left\{J\left(S_{i}^{x} S_{j}^{x}+S_{i}^{y} S_{j}^{y}+\Delta S_{i}^{z} S_{j}^{z}\right)\right.\\ &+2 J_{\pm \pm}\left[\left(S_{i}^{x} S_{j}^{x}-S_{i}^{y} S_{j}^{y}\right) \tilde{c}_{\alpha}-\left(S_{i}^{x} S_{j}^{y}+S_{i}^{y} S_{j}^{x}\right) \tilde{s}_{\alpha}\right] \\ &\left.+J_{z \pm}\left[\left(S_{i}^{y} S_{j}^{z}+S_{i}^{z} S_{j}^{y}\right) \tilde{c}_{\alpha}-\left(S_{i}^{x} S_{j}^{z}+S_{i}^{z} S_{j}^{x}\right) \tilde{s}_{\alpha}\right]\right\}
\end{split}
\label{eq:functionJnn}
\end{align}
where $\tilde{c}(\tilde{s})_{\alpha} = \cos(\sin)\tilde{\phi}_\alpha$ , and $\tilde{\phi}_\alpha = \{0, 2\pi/3, -2\pi/3 \}$ are the bond angles with respect to the $x$-axis. The first term of Eq.(1) is the standard XXZ model and is invariant under the global spin rotation around the $z$-axis. The $J_{\pm\pm}$ and $J_{z\pm}$ terms define the bond dependent anisotropic interactions caused by the strong SOC, and sometimes are called the pseudo-dipolar terms~\cite{Iaconis2018}.

\begin{figure}[!tp]
\centering
\includegraphics[width =0.48\textwidth]{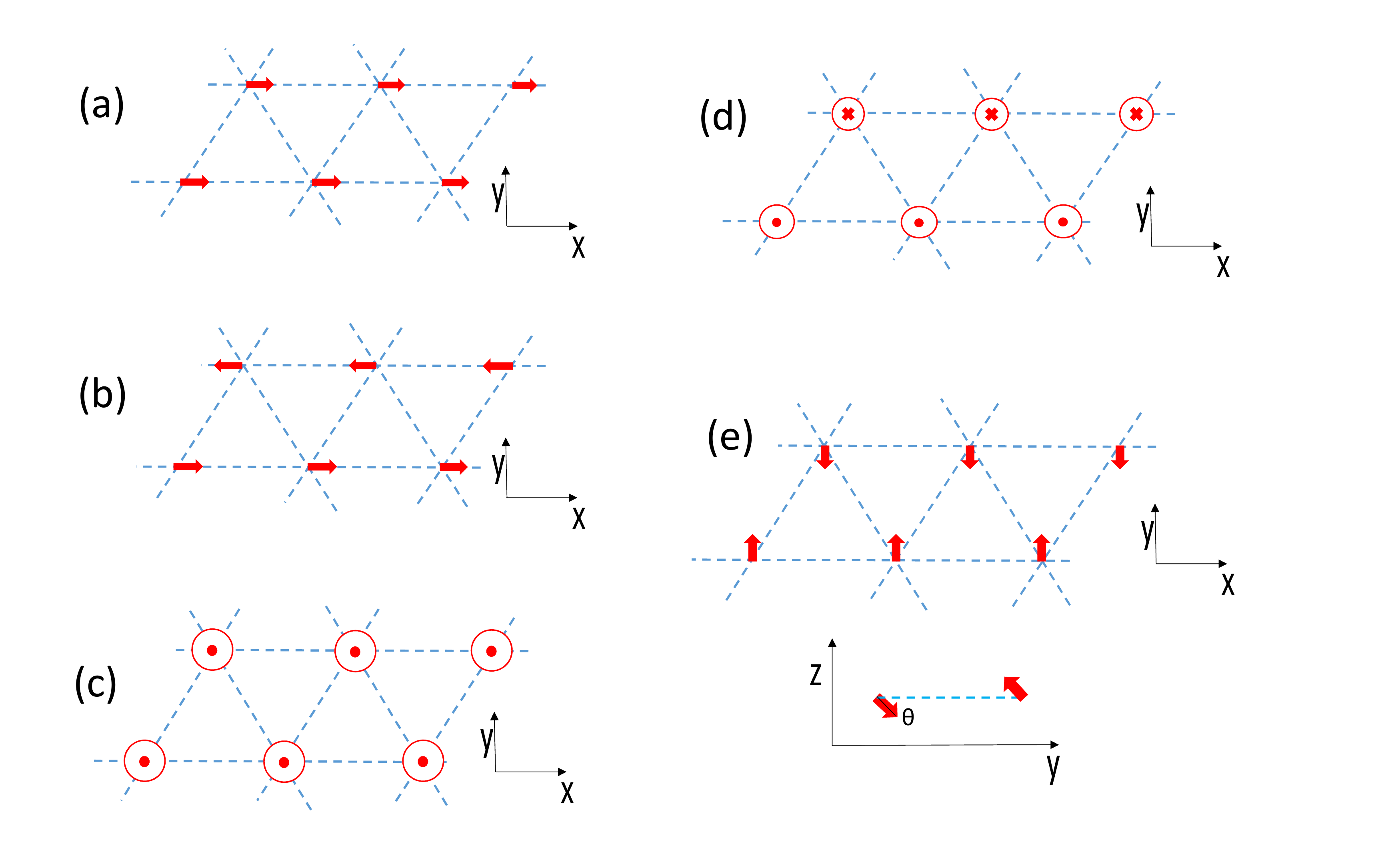}
\caption{(Color online) The spin configurations used to fit the exchange parameters:
(a) All spins are parallel along the $x$ axis (FMx); (b) The spins have a stripe order along the $x$ axis ($x$-stripe); (c) All spins are parallel along the $z$ axis (FMz); (d) The spins have a stripe order along the $z$ axis ($z$-stripe); (e) The spins have a stripe order in the $yz$ plane, with $\theta=\pi / 4$ ($yz$-stripe). }
\label{fig:spin structure}
\end{figure}

To obtain the $J$, $\Delta$,  $J_{\pm\pm}$  and $J_{z\pm}$ parameters, we fit the model to the total energies of five
spin configurations, including  FM$_x$, FM$_z$, $x$-stripe, $z$-stripe, $yz$-stripe states, which are schematically shown in Fig.~\ref{fig:spin structure}. The per-site energies [in units of $S(S+1)$] of these chosen classical spin configurations are as follows:
%~\cite{Witczak2014}:
%
\begin{eqnarray}
E_{{\rm FM}_x} &=& 3J ,\nonumber\\
E_{{\rm FM}_z} &=& 3J\Delta , \nonumber\\
E_{x{\rm-stripe}} &=& -J + 4J_{\pm\pm}, \nonumber\\
E_{z{\rm-stripe}} &=& -J\Delta , \nonumber\\
E_{yz{\rm-stripe}} &=& -\tilde{J}_c - \Delta J - \sqrt{4J_{z\pm}^2 + \tilde{J}_c^2},
\end{eqnarray}
where $\tilde{J}_c = [J(1-\Delta)+4J_{\pm\pm}]/2 $.

To accommodate the above magnetic states, we adopt a supercell containing the 2$\times$2$\times$1 conventional unit cell
of Fig. 1(b). A 9$\times$9$\times$4 $k$-point mesh is used to ensure the convergence of the total energies.
%Due to the strong geometric frustration, it is very difficult to converge the total energies with SOC turned on
%in the self-consistent calculations.
%To calculate the total energies of different magnetic states, we first do the self-consistent calculation for the
%given (non-collinear) magnetic structure without turn on the SOC. We then turn the SOC and perform non-self-consistent calculations
%using the charge densities from the non-SOC calculations.

\begin{table}[!htp]
\caption{The total energies (in eV) of the five spin configurations shown in Fig.\ref{fig:spin structure}.}
\begin{tabular}{ c   c   c   c   c   c}
\hline
\hline
   &FMz  &FMx  &z-stripe &x-stripe &yz-stripe   \\
\hline
NaYbS$_2$   &-201.123  &-201.121    &-201.234  &-201.233  &-201.233  \\
NaYbO$_2$   &-264.303  &-264.302    &-264.317  &-264.315  &-264.316  \\
\hline
\hline
\end{tabular}
\end{table}

\begin{table}[!htp]
\caption{The exchange parameters fitted from total energies of different spin configurations.}
\begin{tabular}{c  c  c  c  c  c  c  c  }
\hline
\hline
  &$J$ (K) &$\Delta$  &$J_{\pm\pm}/J$ &$J_{z\pm}/J$ &$\theta_{CWx}$ &$\theta_{CWz}$ &$\theta^{\rm exp}_{CW}$ \\

NaYbS$_2$ &36.660  &0.980  &1.80$\%$ &2.96$\%$ &-54.99 &-53.87 &-63.74\cite{liu2018} \\
       &       &        &           &          &     &      &  -65 \cite{Baenitz2018} \\
NaYbO$_2$ &5.039  &0.889    &12.81$\%$ &21.1$\%$ &-7.56 &-6.72 & -6 \cite{Ranjith2019} \\
       &       &        &           &          &     &      &  -5.64\cite{leiding2019}  \\
\hline
\hline
\end{tabular}
\end{table}

The calculated total energies of NaYbS$_2$ and  NaYbO$_2$ with the five spin configurations
are listed in Table I, and the fitted exchange parameters
are listed in Table II. The Curie-Weiss (CW) temperatures $\theta_{CW}$ are
estimated as $\theta_{{\rm CW}x}$=-${3 \over 2}$$J$,
whereas $\theta_{{\rm CW}z}$=-${3 \over 2}$$\Delta$$J$\cite{liyuesheng2015-prl,leiding2019}.
Experimentally, $\theta_{\rm CW}$ is fitted from the magnetic susceptibility $1/\chi(T)$ via the Curie-Weiss law.
We mark that $\theta_{\rm CW}$ depend strongly on the fitting temperature. For example, Curie-Weiss temperature is $\theta_{CW}$= -65 K for NaYbS2 when fitting $1/\chi(T)$ below 80K~\cite{Baenitz2018}. However,
$\theta_\perp$=-13.5 K which fitted below 10 K and $\theta_\parallel$=-4.5 K which fitted below 5 K~\cite{Baenitz2018}.
Here, $\perp$ and $\parallel$ refer to that a small magnetic field is applied perpendicular or parallel to the $c$ axis, respectively, when measuring $\chi(T)$.
We compare the calculated $\theta_{\rm CW}$ to the experimental results fitted at higher temperatures.
This is because, at low temperature, the magnetic state tends to be in a strongly correlated (highly entangled) state, whereas at a higher temperature, the spins are more like the classical spin states,
which are more appropriate for the mean-field description.

The calculated $\theta_{\rm CW}$ of NaYbS$_2$ is about 54 K, and that of NaYbO$_2$ is about
6 K, both are in very good agreement with experimental results
\cite{liu2018,Baenitz2018,Ranjith2019,leiding2019,Bordelon2019}.
At first glance, it is somehow surprising, that the magnetic exchange interactions in NaYbS$_2$ are even stronger than those of  NaYbO$_2$, given that the lattice constants of NaYbS$_2$ ($a$=3.901\AA) are larger than those of NaYbO$_2$ ($a$=3.346\AA), due to the larger ion radii of S ions.
However, as seen from the PDOS shown in Fig. 2(b) and Fig. 2(d), the Yb 4$f$ electrons hybridize more strongly
with S 3$p$ electrons than with O 2$p$ electrons, which leads to larger super-exchange interactions.

The SOC interactions introduce the anisotropic magnetic interactions. From Table II, we see that the anisotropy is rather small in NaYbS$_2$, as $\Delta \approx$0.98 (where $\Delta$=1 is the isotropic case).
The anisotropic exchanges $J_{\pm\pm}/J \approx$ 0.018 and $J_{z\pm}/J \approx$ 0.03 are also quite small. NaYbO$_2$ shows somehow stronger anisotropy, with $\Delta \approx $0.89,  $J_{\pm\pm}/J \approx$ 0.128 and $J_{z\pm}/J \approx$ 0.211.

\subsection{Magnetic phase diagram of classical spin model}

The phase diagrams of the classical spin model in Eq.~\ref{eq:functionJnn}
has been studied via spin-wave \cite{Maksimov2019}
and classical Monte Carlo method \cite{liu_semiclassical_2016,liyaodong2016,Parker2018,zhangmeng_unpublished}.
In the vicinity of isotropic region, i.e., $\Delta \approx$1, $J_{\pm\pm}/J \approx$0 and $J_{z\pm}/J \approx$0,
the ground state of model Eq.~1 is a planar 120\degree~ N\'{e}el state.
For $J_{\pm\pm}/J \lesssim$-0.15, the system has a stripe-$x$ order, in which the spins lie within
the $x$-$y$ plane \cite{Maksimov2019},
whereas for $J_{\pm\pm}/J \gtrsim$-0.15, the system is in the stripe-$yz$ order,
where spins are partially out of the $x$-$y$ plane \cite{Maksimov2019}.
Between the stripe phases and the planar 120\degree phase, there are also so-called multi-$Q$ phase,
where the spins are incommensurate and ordered at multiple $Q$ vectors~\cite{liu_semiclassical_2016}.
In the Heisenberg limit, spin-wave results suggest that the multi-$Q$ state is similar
to the $Z_2$ vortex state which has been found in the triangular Kitaev-Heisenberg model \cite{rousochatzakis_kitaev_2016,becker_spin-orbit_2015}.

The quantum spin model has been studied using DMRG methods~\cite{zhu_topography_2018},
and the results suggest that there exists a QSL phase instead of multi-$Q$ phase within the region $J_{z\pm}\simeq[0.27,0.45]$ and $J_{\pm\pm}\simeq[-0.17,0.1]$ in the isotropic limit $\Delta$=1.
The exact diagonalizations \cite{wu_exact_2020} of small clusters of 12 - 32 sites also suggest that there is a spin liquid region, but the spin structure factors are different from the DMRG results.
However, very recently projected entangled pair states (PEPS)~\cite{verstraete04,Verstraete2008,Liuwy2018}
calculations show that there is no QSL in the phase-diagram\cite{zhangmeng_unpublished}.
The nature of the quantum phase in this region is still under debate.

Nevertheless, the calculated exchange interaction parameters for  NaYbS$_2$ and  NaYbO$_2$ are far away from the
DMRG calculated QSL region. In fact,
the parameters calculated for NaYbS$_2$ is very close to the isotopic region, and
the ground state is the planar 120\degree~ N\'{e}el state, whereas
the ground state of NaYbO$_2$ is of the stripe-$yz$ order.

\subsubsection{Nearest Neighbor Model}

To investigate the magnetic phase transitions of NaYbO$_2$, and NaYbS$_2$,
we perform replica-exchange Monte Carlo (MC) simulations~\cite{cao2009first} of the classical spin model of  Eq.~\ref{eq:functionJnn}, using the exchange parameters
obtained from first-principles calculations Sec.\ref{sec:exchange_parameters}.
The simulations are performed on the $L$$\times$$L$ lattices, where $L$=36, 48, 60, 96, and 120.

\begin{figure}
    \centering
    \includegraphics[width=0.4\textwidth]{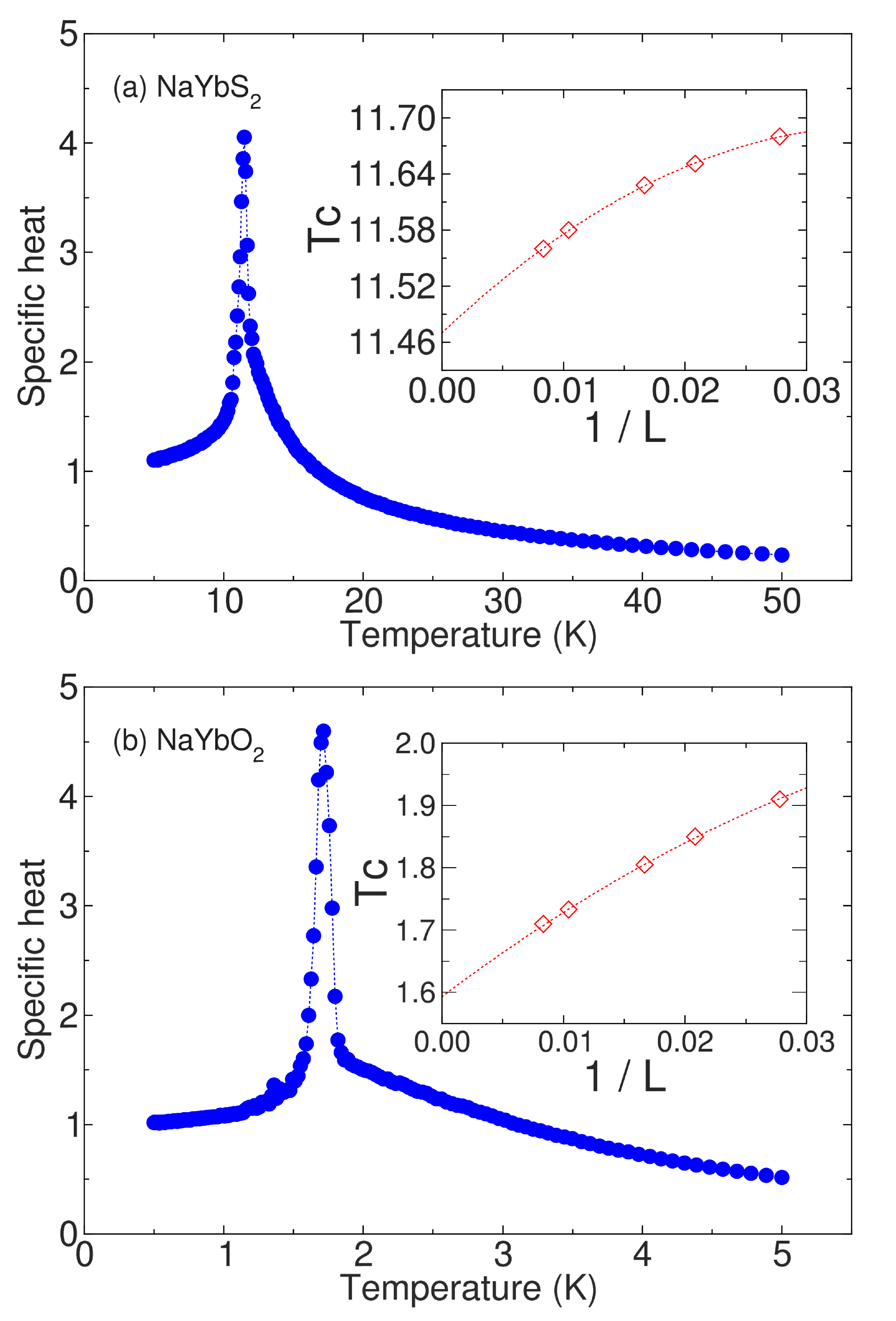}
    \caption{(Color online) The simulated specific heat (blue dots) as functions of temperature for
    (a) NaYbS$_{2}$ and (b) NaYbO$_{2}$ by using Hamiltonian Eq.~\ref{eq:functionJnn}.
     The simulations are performed on a 120$\times$120 lattice.
     The inserts depict the transition temperatures $T_c$ as functions of $1/L$.}
    \label{fig:CvNn}
\end{figure}

Figure~\ref{fig:CvNn}(a),(b) depict the specific heats as functions of temperature for NaYbS$_2$ and NaYbO$_2$
respectively on a 120$\times$120 lattice.
For NaYbO$_2$, the specific heat shows a sharp peak near the temperature of 1.7 K.
In the insert of the figure, we plot the transition temperature calculated on different lattice sizes. By finite-size scaling, the transition temperature in the thermodynamic limit is about 1.6 K.
The magnetic transition for  NaYbS$_2$ is about 11.6 K for 120$\times$120 lattice and about 11.5 K in the thermodynamic limit as shown in Fig.~\ref{fig:CvNn}(b).
These transition temperatures are much too high compared to the experimental results,
where no magnetic transition was observed down to 50 mK for NaYbO$_2$\cite{Bordelon2019}
and 260 mK for NaYbS$_2$ \cite{Baenitz2018}.

%%%%%%%%%%%%%%%%%%%%%%%%%%%%%%%%%%%%%%%%%%%%%%%%%%%%%%%%%%%%%%%%%%%%%%%%%%%%%%%%%%%%%%%%%%
%%%%%%%%%%%%%%%%%%%%%%%%%%%%%%%%%%%%%%%%%%%%%%%%%%%%%%%%%%%%%%%%%%%%%%%%%%%%%%%%%%%%%%%%%%
%\red{Discuss the result of the NN model. Transition temperature too high. }

%%%%%%%%%%%%%%%%%%%%%%%%%%%%%%%%%%%%%%%%%%%%%%%%%%%%%%%%%%%%%%%%%%%%%%%%%%%%%%%%%%%%%%%%%%
\begin{figure}
    \centering
    \includegraphics[width=0.4\textwidth]{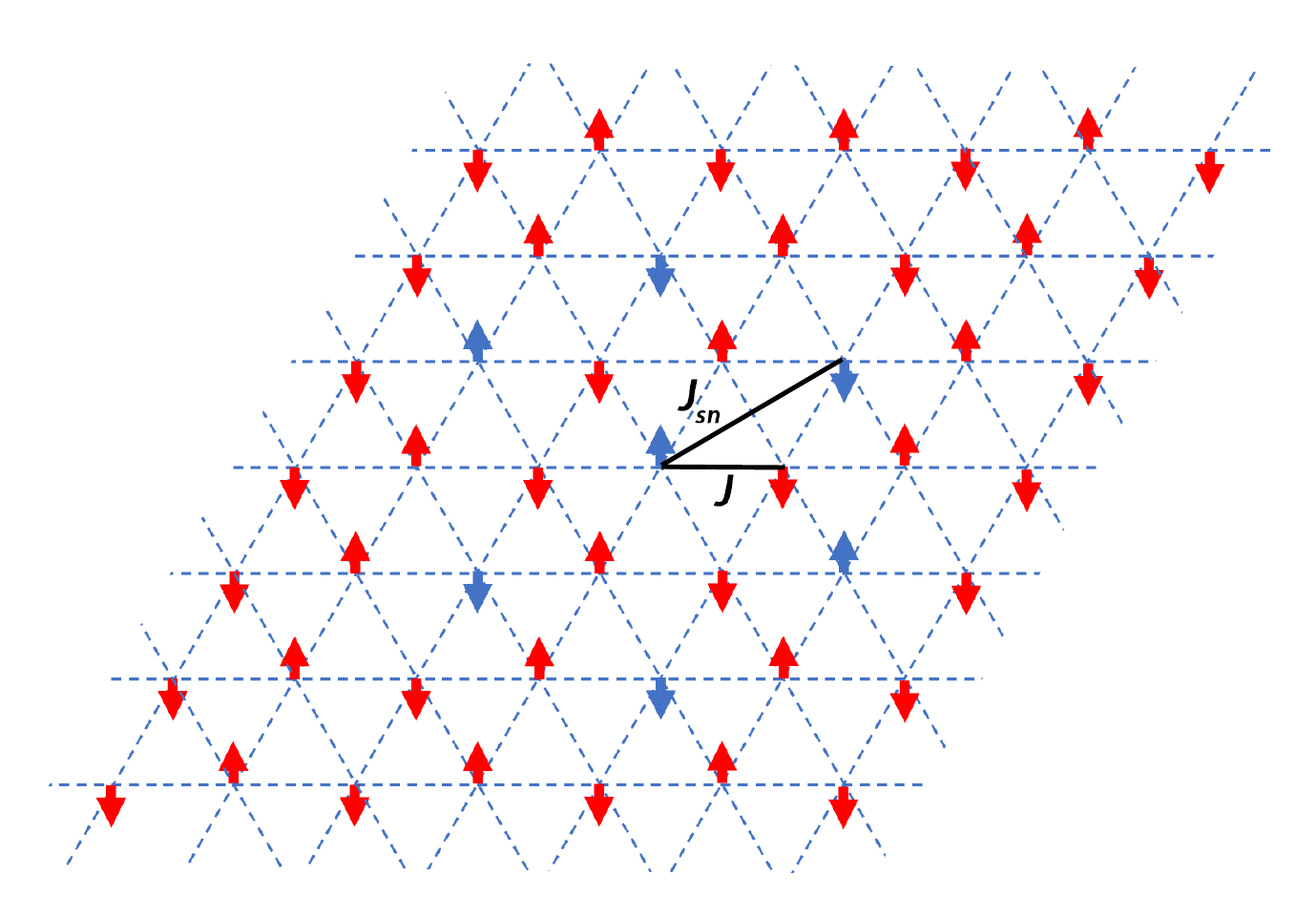}
    \caption{(Color online) The ground state spin configure of NaYbO$_{2}$ with only
   the NN interactions is of a $yz$-stripe order. A spin and all its NNN spins are shown in blue.
   $J$ and $J_{sn}$ are the NN and NNN exchange interactions respectively.}
    \label{fig:NaYbO2Gs}
\end{figure}

\subsubsection{Second Neighbor Model}

Given the above MC simulation results,
we conclude that the experimentally observed magnetic disorder states of NaYbO$_2$ and NaYbS$_2$
are unlikely to be described by the nearest-neighbor (NN) interaction models.
To understand the experimental results, we try to include the next nearest neighboring (NNN) interactions in the MC simulations.
We consider the simplest NNN interactions, which take the form of,
\begin{equation}
\begin{gathered}
    \mathcal{H}_{\rm NNN} = \sum_{\langle\langle i, j \rangle\rangle }J_{sn}(S^{x}_{i}S^{x}_{j}+S^{y}_{i}S^{y}_{j}+\Delta S^{z}_{i}S^{z}_{j})
\end{gathered}
\label{eq:functionJsn}
\end{equation}
Since the first-principles calculations of the NNN exchanges, which require very large supercells,
are extremely difficult for the geometrically frustrated materials NaYbO$_2$ and NaYbS$_2$,  we take the NNN exchange interaction $J_{sn}$ as a parameter, which varies in the range of -0.2$J$ to 0.2$J$.

\begin{figure}[!tb]
    \centering
    \includegraphics[width=0.4\textwidth]{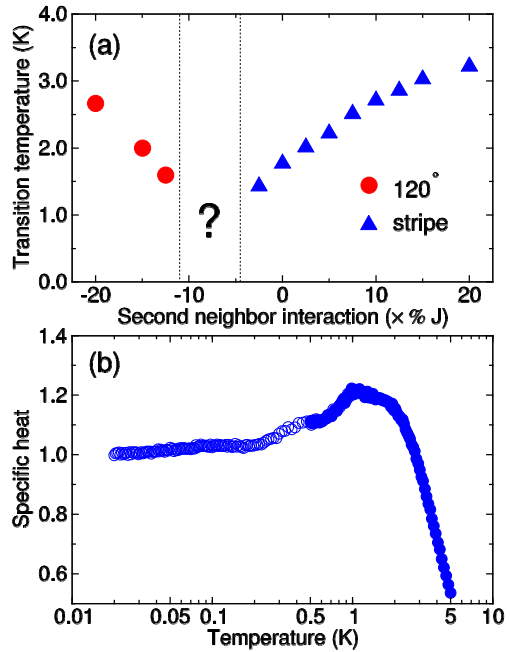}
    \caption{(Color online) (a) The transition temperature $T_c$ as a function of $J_{sn}$.
    For $J_{sn}$$\le$-0.125$J$, the ground state is the  planar 120\degree~ N\'{e}el state (red circles),
    whereas for $J_{sn}$$>$-0.05$J$, the ground state is of $yz$-stripe order (blue triangles).
    For -0.1$J$$\leq$$J_{sn}$$\leq$ -0.05$J$, the ground state is a $Z_2$ vortex phase, with no sharp
    phase transitions. (b) The specific heat as function of temperature for $J_{sn}$= -0.05$J$. }
     \label{fig:NaYbO2Jsn}
\end{figure}

Here, we focus on the results of NaYbO$_{2}$ in the following discussions.
The ground state spin configuration for NaYbO$_{2}$ with NN exchange interactions
is shown in Fig.~\ref{fig:NaYbO2Gs}, which is in a $yz$-stripe order.
A spin and all its NNN spins are shown in blue.
The transition temperature as a function of NNN exchange interaction $J_{sn}$ is shown in Fig.~\ref{fig:NaYbO2Jsn}.
The simulations are carried out on a 48$\times$48 lattice.
At $J_{sn}$=0, the magnetic ground state is in a $yz$-stripe order. When increasing $J_{sn}$ from 0 to 0.2$J$, the magnetic ground state does not change, whereas the transition temperature gradually increases with
the increasing of $J_{sn}$, and reaches about 3 K at $J_{sn}$=0.2$J$.
To understand the results, we note that for each spin on the lattice,
there are six NNN spins around it, as shown in Fig~\ref{fig:NaYbO2Gs}.
In the $yz$-stripe phase, two-third of NNN spins are antiparallel to the central spin
and the other one-third of spins are parallel to it.
When a positive $J_{sn}$ in Eq.~\ref{eq:functionJsn} is used,
the NNN interactions further stabilize the magnetic order and therefore increase the transition temperature.

When a negative $J_{sn}$ is added, the NNN interactions add more frustration to the $yz$-stripe order
which will decrease the Curie temperature as shown in  Fig.~\ref{fig:NaYbO2Jsn}.
For $J_{sn}$$\le$-0.125$J$, the ground state becomes to the  planar 120\degree~ N\'{e}el state,
and the transition temperature raises with the decreasing $J_{sn}$.

%%%%%%%%%%%%%%%%%%%%%%%%%%%%%%%%%%%%%%%%%%%%%%%%%%%%%%%%%%%%%%%%%%%%%%%%%%%%%%%%%%%%%%%%%
\begin{figure}
    \centering
    \includegraphics[width=0.5\textwidth]{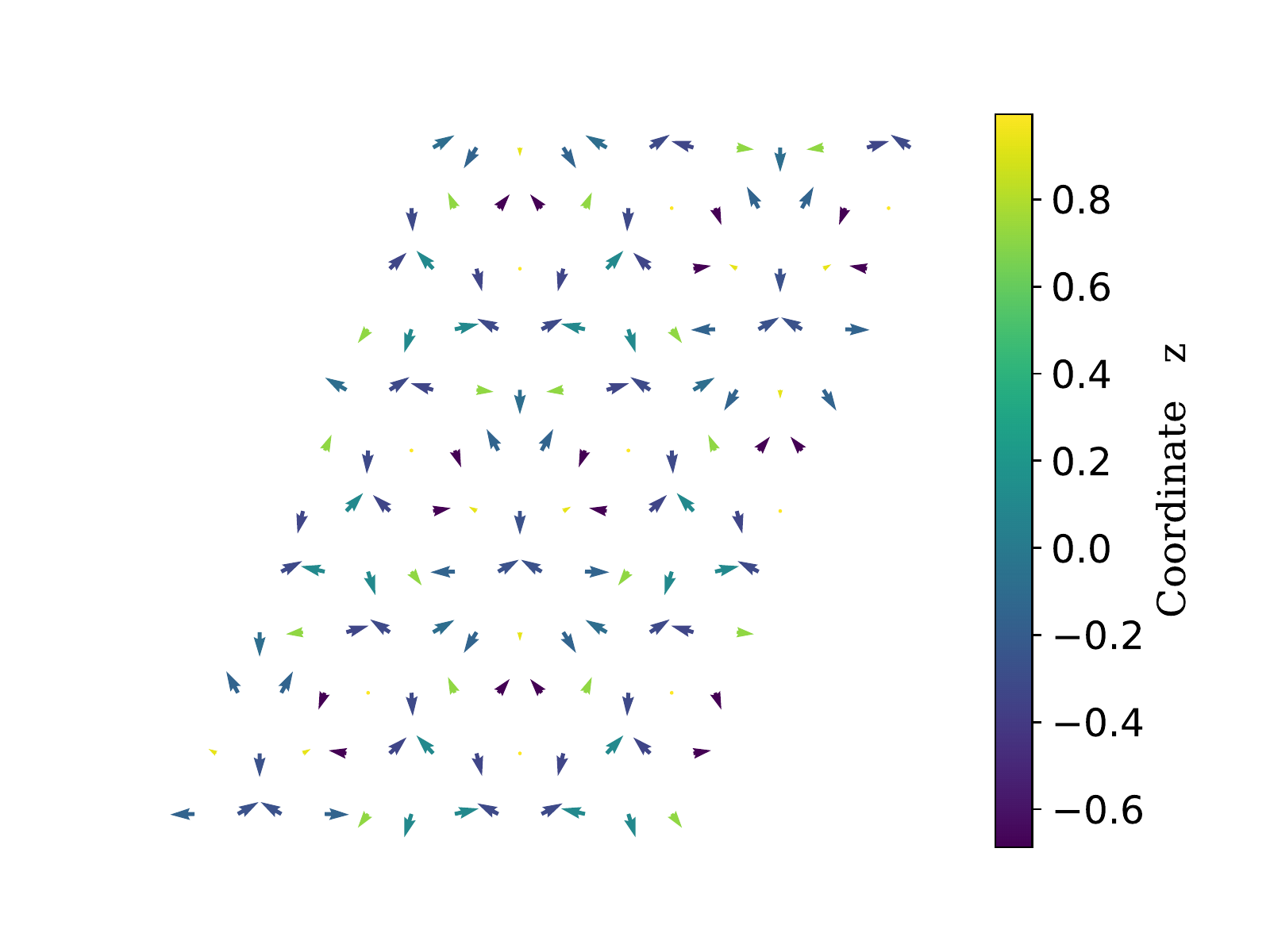}
    \caption{(Color online) The ground state spin configure of the $Z_2$ vortex phase.
    The arrows show the $x$, $y$ components of the spins,
    whereas the color marks the $z$ component. }
    \label{fig:vortexGs}
\end{figure}
%%%%%%%%%%%%%%%%%%%%%%%%%%%%%%%%%%%%%%%%%%%%%%%%%%%%%%%%%%%%%%%%%%%%%%%%%%%%%%%%%%%%%%%%%%

Remarkably, for -0.1$J$$\leq$$J_{sn}$$\leq$ -0.05$J$, we find no obvious phase transition down to the
lowest simulation temperature of 20 mK.
The specific heat as a function of temperature for $J_{sn}$=-0.05$J$ has a very broad peak as shown in Fig.~\ref{fig:NaYbO2Jsn}(b), which is very similar to the $C_v$ curve for NaYbO$_2$ at $H$=0~\cite{Bordelon2019}.
In the experiments, $C_v$ approach zero as temperature approaching zero\cite{Bordelon2019}, while $C_v$ approach a finite value
in our simulation. This might because we use a classical model, instead of a quantum model.
The ground state spin configuration is shown in Fig~\ref{fig:vortexGs}, which can be identified as a $Z_2$ vortex phase
\cite{rousochatzakis_kitaev_2016,becker_spin-orbit_2015,zhangmeng_unpublished}.
%\red{For vortex ground state spin configures, the direction of spin forms different circles, those circles are periodic in long-range.}
The broad $C_v$ curve is due to the Kosterlitz-Thouless (KT) melting of the $Z_2$ vortex \cite{kawamura84}.
We note that recently the KT transition has also been proposed by Li et. al. for TmMgGaO$_4$~\cite{Lihan2020}.
Whether the $Z_2$ ground state will melt at zero temperature due to quantum fluctuation resulting
in a QSL state, requires further studies.

\section{Summary}

We calculate the magnetic interactions in NaYbO$_2$ and NaYbS$_2$ via the first-principles method.
The calculated Curie-Weiss temperatures are in good agreement with experiments.
We then perform classical MC simulations of the finite-temperature phase diagram of the two compounds using the
extracted exchange parameters. We find that if only the nearest neighbor interactions are considered, the ground states are a stripe and a planar 120\degree~ N\'{e}el state for NaYbO$_2$ and
NaYbS$_2$, respectively. The simulated transition temperatures are much higher than the lowest experimental temperatures, where no magnetic ordering was observed. These results suggest that the experimentally observed magnetic disorder states of NaYbO$_2$ and NaYbS$_2$ are unlikely to be described by the nearest-neighbor interaction models.
We show by adding suitable second neighboring interactions, the classical magnetic ground state of NaYbO$_2$ becomes to the $Z_2$ vortex phase, and the simulated specific heat $C_v$ has a very broad peak, which is due to the KT melting of the $Z_2$ vortex. Whether the $Z_2$ ground state will melt due to quantum fluctuation at zero temperature resulting in a QSL state
is an interesting topic for future studies.

\acknowledgements
This work is funded by the Chinese National Science Foundation Grant number 11774327.
The numerical calculations were done on the USTC HPC facilities.

%
%\bibliographystyle{apsrev}
%\bibliography{NaYbO2-ref}

\begin{thebibliography}{43}
\expandafter\ifx\csname natexlab\endcsname\relax\def\natexlab#1{#1}\fi
\expandafter\ifx\csname bibnamefont\endcsname\relax
  \def\bibnamefont#1{#1}\fi
\expandafter\ifx\csname bibfnamefont\endcsname\relax
  \def\bibfnamefont#1{#1}\fi
\expandafter\ifx\csname citenamefont\endcsname\relax
  \def\citenamefont#1{#1}\fi
\expandafter\ifx\csname url\endcsname\relax
  \def\url#1{\texttt{#1}}\fi
\expandafter\ifx\csname urlprefix\endcsname\relax\def\urlprefix{URL }\fi
\providecommand{\bibinfo}[2]{#2}
\providecommand{\eprint}[2][]{\url{#2}}

\bibitem[{\citenamefont{Anderson}(1973)}]{Anderson1973}
\bibinfo{author}{\bibfnamefont{P.~W.} \bibnamefont{Anderson}},
  \bibinfo{journal}{Mater. Res. Bull.} \textbf{\bibinfo{volume}{8}},
  \bibinfo{pages}{153} (\bibinfo{year}{1973}).

\bibitem[{\citenamefont{Anderson}(1987)}]{Anderson1987}
\bibinfo{author}{\bibfnamefont{P.~W.} \bibnamefont{Anderson}},
  \bibinfo{journal}{Science} \textbf{\bibinfo{volume}{235}},
  \bibinfo{pages}{1196} (\bibinfo{year}{1987}).

\bibitem[{\citenamefont{Balents}(2010)}]{Balents2010}
\bibinfo{author}{\bibfnamefont{L.}~\bibnamefont{Balents}},
  \bibinfo{journal}{Nature} \textbf{\bibinfo{volume}{464}},
  \bibinfo{pages}{199} (\bibinfo{year}{2010}).

\bibitem[{\citenamefont{Savary and Balents}(2017)}]{Savary2017}
\bibinfo{author}{\bibfnamefont{L.}~\bibnamefont{Savary}} \bibnamefont{and}
  \bibinfo{author}{\bibfnamefont{L.}~\bibnamefont{Balents}},
  \bibinfo{journal}{Rep. Prog. Phys.} \textbf{\bibinfo{volume}{80}},
  \bibinfo{pages}{016502} (\bibinfo{year}{2017}).

\bibitem[{\citenamefont{Lacroix et~al.}(1988)\citenamefont{Lacroix, Mendels,
  and Mila}}]{frustrated_magnetism_book}
\bibinfo{editor}{\bibfnamefont{C.}~\bibnamefont{Lacroix}},
  \bibinfo{editor}{\bibfnamefont{P.}~\bibnamefont{Mendels}}, \bibnamefont{and}
  \bibinfo{editor}{\bibfnamefont{F.}~\bibnamefont{Mila}}, eds.,
  \emph{\bibinfo{title}{Introduction to Frustrated Magnetism}}
  (\bibinfo{publisher}{Springer-Verlag}, \bibinfo{address}{Berlin Heidelberg},
  \bibinfo{year}{1988}).

\bibitem[{\citenamefont{Li et~al.}(2015)\citenamefont{Li, Chen, Tong, Pi, Liu,
  Yang, Wang, and Zhang}}]{liyuesheng2015-prl}
\bibinfo{author}{\bibfnamefont{Y.}~\bibnamefont{Li}},
  \bibinfo{author}{\bibfnamefont{G.}~\bibnamefont{Chen}},
  \bibinfo{author}{\bibfnamefont{W.}~\bibnamefont{Tong}},
  \bibinfo{author}{\bibfnamefont{L.}~\bibnamefont{Pi}},
  \bibinfo{author}{\bibfnamefont{J.}~\bibnamefont{Liu}},
  \bibinfo{author}{\bibfnamefont{Z.}~\bibnamefont{Yang}},
  \bibinfo{author}{\bibfnamefont{X.}~\bibnamefont{Wang}}, \bibnamefont{and}
  \bibinfo{author}{\bibfnamefont{Q.}~\bibnamefont{Zhang}},
  \bibinfo{journal}{Phys. Rev. Lett.} \textbf{\bibinfo{volume}{115}},
  \bibinfo{pages}{167203} (\bibinfo{year}{2015}).

\bibitem[{\citenamefont{Li et~al.}(2016{\natexlab{a}})\citenamefont{Li, Wang,
  and Chen}}]{liyaodong2016}
\bibinfo{author}{\bibfnamefont{Y.-D.} \bibnamefont{Li}},
  \bibinfo{author}{\bibfnamefont{X.}~\bibnamefont{Wang}}, \bibnamefont{and}
  \bibinfo{author}{\bibfnamefont{G.}~\bibnamefont{Chen}},
  \bibinfo{journal}{Phys. Rev. B} \textbf{\bibinfo{volume}{94}},
  \bibinfo{pages}{035107} (\bibinfo{year}{2016}{\natexlab{a}}).

\bibitem[{\citenamefont{Li et~al.}(2016{\natexlab{b}})\citenamefont{Li, Adroja,
  Biswas, Baker, Zhang, Liu, Tsirlin, Gegenwart, and Zhang}}]{Li2016}
\bibinfo{author}{\bibfnamefont{Y.}~\bibnamefont{Li}},
  \bibinfo{author}{\bibfnamefont{D.}~\bibnamefont{Adroja}},
  \bibinfo{author}{\bibfnamefont{P.~K.} \bibnamefont{Biswas}},
  \bibinfo{author}{\bibfnamefont{P.~J.} \bibnamefont{Baker}},
  \bibinfo{author}{\bibfnamefont{Q.}~\bibnamefont{Zhang}},
  \bibinfo{author}{\bibfnamefont{J.}~\bibnamefont{Liu}},
  \bibinfo{author}{\bibfnamefont{A.~A.} \bibnamefont{Tsirlin}},
  \bibinfo{author}{\bibfnamefont{P.}~\bibnamefont{Gegenwart}},
  \bibnamefont{and} \bibinfo{author}{\bibfnamefont{Q.}~\bibnamefont{Zhang}},
  \bibinfo{journal}{Phys. Rev. Lett.} \textbf{\bibinfo{volume}{117}},
  \bibinfo{pages}{097201} (\bibinfo{year}{2016}{\natexlab{b}}).

\bibitem[{\citenamefont{Shen et~al.}(2016)\citenamefont{Shen, Li, Wo, Li, Shen,
  Pan, Wang, Walker, Steffens, Boehm et~al.}}]{Shen2016}
\bibinfo{author}{\bibfnamefont{Y.}~\bibnamefont{Shen}},
  \bibinfo{author}{\bibfnamefont{Y.-D.} \bibnamefont{Li}},
  \bibinfo{author}{\bibfnamefont{H.}~\bibnamefont{Wo}},
  \bibinfo{author}{\bibfnamefont{Y.}~\bibnamefont{Li}},
  \bibinfo{author}{\bibfnamefont{S.}~\bibnamefont{Shen}},
  \bibinfo{author}{\bibfnamefont{B.}~\bibnamefont{Pan}},
  \bibinfo{author}{\bibfnamefont{Q.}~\bibnamefont{Wang}},
  \bibinfo{author}{\bibfnamefont{H.~C.} \bibnamefont{Walker}},
  \bibinfo{author}{\bibfnamefont{P.}~\bibnamefont{Steffens}},
  \bibinfo{author}{\bibfnamefont{M.}~\bibnamefont{Boehm}},
  \bibnamefont{et~al.}, \bibinfo{journal}{Nature}
  \textbf{\bibinfo{volume}{540}}, \bibinfo{pages}{559} (\bibinfo{year}{2016}).

\bibitem[{\citenamefont{Paddison et~al.}(2017)\citenamefont{Paddison, Daum,
  Dun, Ehlers, Liu, Stone, Zhou, and Mourigal}}]{Paddison2017}
\bibinfo{author}{\bibfnamefont{J.~A.~M.} \bibnamefont{Paddison}},
  \bibinfo{author}{\bibfnamefont{M.}~\bibnamefont{Daum}},
  \bibinfo{author}{\bibfnamefont{Z.}~\bibnamefont{Dun}},
  \bibinfo{author}{\bibfnamefont{G.}~\bibnamefont{Ehlers}},
  \bibinfo{author}{\bibfnamefont{Y.}~\bibnamefont{Liu}},
  \bibinfo{author}{\bibfnamefont{M.}~\bibnamefont{Stone}},
  \bibinfo{author}{\bibfnamefont{H.}~\bibnamefont{Zhou}}, \bibnamefont{and}
  \bibinfo{author}{\bibfnamefont{M.}~\bibnamefont{Mourigal}},
  \bibinfo{journal}{Nat. Phys.} \textbf{\bibinfo{volume}{13}},
  \bibinfo{pages}{117} (\bibinfo{year}{2017}).

\bibitem[{\citenamefont{Zhu et~al.}(2017)\citenamefont{Zhu, Maksimov, White,
  and Chernyshev}}]{Zhu2017}
\bibinfo{author}{\bibfnamefont{Z.}~\bibnamefont{Zhu}},
  \bibinfo{author}{\bibfnamefont{P.~A.} \bibnamefont{Maksimov}},
  \bibinfo{author}{\bibfnamefont{S.~R.} \bibnamefont{White}}, \bibnamefont{and}
  \bibinfo{author}{\bibfnamefont{A.~L.} \bibnamefont{Chernyshev}},
  \bibinfo{journal}{Phys. Rev. Lett.} \textbf{\bibinfo{volume}{119}},
  \bibinfo{pages}{157201} (\bibinfo{year}{2017}).

\bibitem[{\citenamefont{Parker and Balents}(2018)}]{Parker2018}
\bibinfo{author}{\bibfnamefont{E.}~\bibnamefont{Parker}} \bibnamefont{and}
  \bibinfo{author}{\bibfnamefont{L.}~\bibnamefont{Balents}},
  \bibinfo{journal}{Phys. Rev. B} \textbf{\bibinfo{volume}{97}},
  \bibinfo{pages}{184413} (\bibinfo{year}{2018}).

\bibitem[{\citenamefont{Liu et~al.}(2018{\natexlab{a}})\citenamefont{Liu,
  Zhang, Ji, Liu, Li, Wang, Lei, Chen, and Zhang}}]{liu2018}
\bibinfo{author}{\bibfnamefont{W.}~\bibnamefont{Liu}},
  \bibinfo{author}{\bibfnamefont{Z.}~\bibnamefont{Zhang}},
  \bibinfo{author}{\bibfnamefont{J.}~\bibnamefont{Ji}},
  \bibinfo{author}{\bibfnamefont{Y.}~\bibnamefont{Liu}},
  \bibinfo{author}{\bibfnamefont{J.}~\bibnamefont{Li}},
  \bibinfo{author}{\bibfnamefont{X.}~\bibnamefont{Wang}},
  \bibinfo{author}{\bibfnamefont{H.}~\bibnamefont{Lei}},
  \bibinfo{author}{\bibfnamefont{G.}~\bibnamefont{Chen}}, \bibnamefont{and}
  \bibinfo{author}{\bibfnamefont{Q.}~\bibnamefont{Zhang}},
  \bibinfo{journal}{Chinese Phys. Lett.} \textbf{\bibinfo{volume}{35}},
  \bibinfo{pages}{117501} (\bibinfo{year}{2018}{\natexlab{a}}).

\bibitem[{\citenamefont{Xing et~al.}(2019)\citenamefont{Xing, Sanjeewa, Kim,
  Stewart, Podlesnyak, and Sefat}}]{xing2019}
\bibinfo{author}{\bibfnamefont{J.}~\bibnamefont{Xing}},
  \bibinfo{author}{\bibfnamefont{L.~D.} \bibnamefont{Sanjeewa}},
  \bibinfo{author}{\bibfnamefont{J.}~\bibnamefont{Kim}},
  \bibinfo{author}{\bibfnamefont{G.~R.} \bibnamefont{Stewart}},
  \bibinfo{author}{\bibfnamefont{A.}~\bibnamefont{Podlesnyak}},
  \bibnamefont{and} \bibinfo{author}{\bibfnamefont{A.~S.} \bibnamefont{Sefat}},
  \bibinfo{journal}{Phys. Rev. B} \textbf{\bibinfo{volume}{100}},
  \bibinfo{pages}{220407(R)} (\bibinfo{year}{2019}).

\bibitem[{\citenamefont{Ranjith et~al.}(2019)\citenamefont{Ranjith, Luther,
  Reimann, Schmidt, Schlender, Sichelschmidt, Yasuoka, Strydom, Skourski,
  Wosnitza et~al.}}]{Ranjith2019}
\bibinfo{author}{\bibfnamefont{K.~M.} \bibnamefont{Ranjith}},
  \bibinfo{author}{\bibfnamefont{S.}~\bibnamefont{Luther}},
  \bibinfo{author}{\bibfnamefont{T.}~\bibnamefont{Reimann}},
  \bibinfo{author}{\bibfnamefont{B.}~\bibnamefont{Schmidt}},
  \bibinfo{author}{\bibfnamefont{P.}~\bibnamefont{Schlender}},
  \bibinfo{author}{\bibfnamefont{J.}~\bibnamefont{Sichelschmidt}},
  \bibinfo{author}{\bibfnamefont{H.}~\bibnamefont{Yasuoka}},
  \bibinfo{author}{\bibfnamefont{A.~M.} \bibnamefont{Strydom}},
  \bibinfo{author}{\bibfnamefont{Y.}~\bibnamefont{Skourski}},
  \bibinfo{author}{\bibfnamefont{J.}~\bibnamefont{Wosnitza}},
  \bibnamefont{et~al.}, \bibinfo{journal}{Phys. Rev. B}
  \textbf{\bibinfo{volume}{100}}, \bibinfo{pages}{224417}
  (\bibinfo{year}{2019}).

\bibitem[{\citenamefont{Sarkar et~al.}(2019)\citenamefont{Sarkar, Schlender,
  Grinenko, Haeussler, Baker, Doert, and Klauss}}]{Sarkar2019}
\bibinfo{author}{\bibfnamefont{R.}~\bibnamefont{Sarkar}},
  \bibinfo{author}{\bibfnamefont{P.}~\bibnamefont{Schlender}},
  \bibinfo{author}{\bibfnamefont{V.}~\bibnamefont{Grinenko}},
  \bibinfo{author}{\bibfnamefont{E.}~\bibnamefont{Haeussler}},
  \bibinfo{author}{\bibfnamefont{P.~J.} \bibnamefont{Baker}},
  \bibinfo{author}{\bibfnamefont{T.}~\bibnamefont{Doert}}, \bibnamefont{and}
  \bibinfo{author}{\bibfnamefont{H.-H.} \bibnamefont{Klauss}},
  \bibinfo{journal}{Phys. Rev. B} \textbf{\bibinfo{volume}{100}},
  \bibinfo{pages}{241116(R)} (\bibinfo{year}{2019}).

\bibitem[{\citenamefont{Bordelon et~al.}(2019)\citenamefont{Bordelon, Kenney,
  Liu, Hogan, Posthuma, Kavand, Lyu, Sherwin, Butch, Brown
  et~al.}}]{Bordelon2019}
\bibinfo{author}{\bibfnamefont{M.~M.} \bibnamefont{Bordelon}},
  \bibinfo{author}{\bibfnamefont{E.}~\bibnamefont{Kenney}},
  \bibinfo{author}{\bibfnamefont{C.}~\bibnamefont{Liu}},
  \bibinfo{author}{\bibfnamefont{T.}~\bibnamefont{Hogan}},
  \bibinfo{author}{\bibfnamefont{L.}~\bibnamefont{Posthuma}},
  \bibinfo{author}{\bibfnamefont{M.}~\bibnamefont{Kavand}},
  \bibinfo{author}{\bibfnamefont{Y.}~\bibnamefont{Lyu}},
  \bibinfo{author}{\bibfnamefont{M.~S.} \bibnamefont{Sherwin}},
  \bibinfo{author}{\bibfnamefont{N.~P.} \bibnamefont{Butch}},
  \bibinfo{author}{\bibfnamefont{C.~M.} \bibnamefont{Brown}},
  \bibnamefont{et~al.}, \bibinfo{journal}{Nat. Phys.}
  \textbf{\bibinfo{volume}{15}}, \bibinfo{pages}{1058} (\bibinfo{year}{2019}).

\bibitem[{\citenamefont{Capriotti et~al.}(1999)\citenamefont{Capriotti,
  Trumper, and Sorella}}]{Capriotti1999}
\bibinfo{author}{\bibfnamefont{L.}~\bibnamefont{Capriotti}},
  \bibinfo{author}{\bibfnamefont{A.~E.} \bibnamefont{Trumper}},
  \bibnamefont{and} \bibinfo{author}{\bibfnamefont{S.}~\bibnamefont{Sorella}},
  \bibinfo{journal}{Phys. Rev. Lett.} \textbf{\bibinfo{volume}{82}},
  \bibinfo{pages}{3899} (\bibinfo{year}{1999}).

\bibitem[{\citenamefont{Perdew et~al.}(1996)\citenamefont{Perdew, Burke, and
  Ernzerhof}}]{pbe-ref}
\bibinfo{author}{\bibfnamefont{J.~P.} \bibnamefont{Perdew}},
  \bibinfo{author}{\bibfnamefont{K.}~\bibnamefont{Burke}}, \bibnamefont{and}
  \bibinfo{author}{\bibfnamefont{M.}~\bibnamefont{Ernzerhof}},
  \bibinfo{journal}{Phys. Rev. Lett.} \textbf{\bibinfo{volume}{77}},
  \bibinfo{pages}{3865} (\bibinfo{year}{1996}).

\bibitem[{\citenamefont{Kresse and Furthm\"uller}(1996)}]{vasp-ref}
\bibinfo{author}{\bibfnamefont{G.}~\bibnamefont{Kresse}} \bibnamefont{and}
  \bibinfo{author}{\bibfnamefont{J.}~\bibnamefont{Furthm\"uller}},
  \bibinfo{journal}{Phys. Rev. B} \textbf{\bibinfo{volume}{54}},
  \bibinfo{pages}{11169} (\bibinfo{year}{1996}).

\bibitem[{\citenamefont{Dudarev et~al.}(1998)\citenamefont{Dudarev, Botton,
  Savrasov, Humphreys, and Sutton}}]{Dudarev1998}
\bibinfo{author}{\bibfnamefont{S.~L.} \bibnamefont{Dudarev}},
  \bibinfo{author}{\bibfnamefont{G.~A.} \bibnamefont{Botton}},
  \bibinfo{author}{\bibfnamefont{S.~Y.} \bibnamefont{Savrasov}},
  \bibinfo{author}{\bibfnamefont{C.~J.} \bibnamefont{Humphreys}},
  \bibnamefont{and} \bibinfo{author}{\bibfnamefont{A.~P.}
  \bibnamefont{Sutton}}, \bibinfo{journal}{Phys. Rev. B}
  \textbf{\bibinfo{volume}{57}}, \bibinfo{pages}{1505} (\bibinfo{year}{1998}).

\bibitem[{\citenamefont{Schleid and Lissner}(1993)}]{NaYbS2-stru-ref}
\bibinfo{author}{\bibfnamefont{T.}~\bibnamefont{Schleid}} \bibnamefont{and}
  \bibinfo{author}{\bibfnamefont{F.}~\bibnamefont{Lissner}},
  \bibinfo{journal}{European journal of solid state and inorganic chemistry}
  \textbf{\bibinfo{volume}{30}}, \bibinfo{pages}{829} (\bibinfo{year}{1993}).

\bibitem[{\citenamefont{Hashimoto et~al.}(2003)\citenamefont{Hashimoto,
  Wakeshima, and Hinatsu}}]{NaYbO2-stru-ref}
\bibinfo{author}{\bibfnamefont{Y.}~\bibnamefont{Hashimoto}},
  \bibinfo{author}{\bibfnamefont{M.}~\bibnamefont{Wakeshima}},
  \bibnamefont{and} \bibinfo{author}{\bibfnamefont{Y.}~\bibnamefont{Hinatsu}},
  \bibinfo{journal}{Journal of Solid State Chemistry}
  \textbf{\bibinfo{volume}{176}}, \bibinfo{pages}{266} (\bibinfo{year}{2003}).

\bibitem[{\citenamefont{Dzyaloshinskii}(1964)}]{dzyaloshinskii64}
\bibinfo{author}{\bibfnamefont{I.~E.} \bibnamefont{Dzyaloshinskii}},
  \bibinfo{journal}{Sov. Phys. JETP} \textbf{\bibinfo{volume}{19}},
  \bibinfo{pages}{960} (\bibinfo{year}{1964}).

\bibitem[{\citenamefont{Moriya}(1960)}]{moriya60}
\bibinfo{author}{\bibfnamefont{T.}~\bibnamefont{Moriya}},
  \bibinfo{journal}{Phys. Rev.} \textbf{\bibinfo{volume}{120}},
  \bibinfo{pages}{91} (\bibinfo{year}{1960}).

\bibitem[{\citenamefont{Baenitz et~al.}(2018)\citenamefont{Baenitz, Schlender,
  Sichelschmidt, Onykiienko, Zangeneh, Ranjith, Sarkar, Hozoi, Walker, Orain
  et~al.}}]{Baenitz2018}
\bibinfo{author}{\bibfnamefont{M.}~\bibnamefont{Baenitz}},
  \bibinfo{author}{\bibfnamefont{P.}~\bibnamefont{Schlender}},
  \bibinfo{author}{\bibfnamefont{J.}~\bibnamefont{Sichelschmidt}},
  \bibinfo{author}{\bibfnamefont{Y.~A.} \bibnamefont{Onykiienko}},
  \bibinfo{author}{\bibfnamefont{Z.}~\bibnamefont{Zangeneh}},
  \bibinfo{author}{\bibfnamefont{K.~M.} \bibnamefont{Ranjith}},
  \bibinfo{author}{\bibfnamefont{R.}~\bibnamefont{Sarkar}},
  \bibinfo{author}{\bibfnamefont{L.}~\bibnamefont{Hozoi}},
  \bibinfo{author}{\bibfnamefont{H.~C.} \bibnamefont{Walker}},
  \bibinfo{author}{\bibfnamefont{J.-C.} \bibnamefont{Orain}},
  \bibnamefont{et~al.}, \bibinfo{journal}{Phys. Rev. B}
  \textbf{\bibinfo{volume}{98}}, \bibinfo{pages}{220409(R)}
  (\bibinfo{year}{2018}).

\bibitem[{\citenamefont{Georges et~al.}(1996)\citenamefont{Georges, Kotliar,
  Krauth, and Rozenberg}}]{DMFT1996}
\bibinfo{author}{\bibfnamefont{A.}~\bibnamefont{Georges}},
  \bibinfo{author}{\bibfnamefont{G.}~\bibnamefont{Kotliar}},
  \bibinfo{author}{\bibfnamefont{W.}~\bibnamefont{Krauth}}, \bibnamefont{and}
  \bibinfo{author}{\bibfnamefont{M.~J.} \bibnamefont{Rozenberg}},
  \bibinfo{journal}{Rev. Mod. Phys.} \textbf{\bibinfo{volume}{68}},
  \bibinfo{pages}{13} (\bibinfo{year}{1996}).

\bibitem[{\citenamefont{Witczak-Krempa
  et~al.}(2014)\citenamefont{Witczak-Krempa, Chen, Kim, and
  Balents}}]{Witczak2014}
\bibinfo{author}{\bibfnamefont{W.}~\bibnamefont{Witczak-Krempa}},
  \bibinfo{author}{\bibfnamefont{G.}~\bibnamefont{Chen}},
  \bibinfo{author}{\bibfnamefont{Y.~B.} \bibnamefont{Kim}}, \bibnamefont{and}
  \bibinfo{author}{\bibfnamefont{L.}~\bibnamefont{Balents}},
  \bibinfo{journal}{Annual Review of Condensed Matter Physics}
  \textbf{\bibinfo{volume}{5}}, \bibinfo{pages}{57} (\bibinfo{year}{2014}).

\bibitem[{\citenamefont{Maksimov et~al.}(2019)\citenamefont{Maksimov, Zhu,
  White, and Chernyshev}}]{Maksimov2019}
\bibinfo{author}{\bibfnamefont{P.~A.} \bibnamefont{Maksimov}},
  \bibinfo{author}{\bibfnamefont{Z.}~\bibnamefont{Zhu}},
  \bibinfo{author}{\bibfnamefont{S.~R.} \bibnamefont{White}}, \bibnamefont{and}
  \bibinfo{author}{\bibfnamefont{A.~L.} \bibnamefont{Chernyshev}},
  \bibinfo{journal}{Phys. Rev. X} \textbf{\bibinfo{volume}{9}},
  \bibinfo{pages}{021017} (\bibinfo{year}{2019}).

\bibitem[{\citenamefont{Iaconis et~al.}(2018)\citenamefont{Iaconis, Liu,
  Halasz, and Balents}}]{Iaconis2018}
\bibinfo{author}{\bibfnamefont{J.}~\bibnamefont{Iaconis}},
  \bibinfo{author}{\bibfnamefont{C.}~\bibnamefont{Liu}},
  \bibinfo{author}{\bibfnamefont{G.~B.} \bibnamefont{Halasz}},
  \bibnamefont{and} \bibinfo{author}{\bibfnamefont{L.}~\bibnamefont{Balents}},
  \bibinfo{journal}{SciPost Phys.} \textbf{\bibinfo{volume}{4}},
  \bibinfo{pages}{3} (\bibinfo{year}{2018}).

\bibitem[{\citenamefont{Ding et~al.}(2019)\citenamefont{Ding, Manuel, Bachus,
  Gru\ss{}ler, Gegenwart, Singleton, Johnson, Walker, Adroja, Hillier
  et~al.}}]{leiding2019}
\bibinfo{author}{\bibfnamefont{L.}~\bibnamefont{Ding}},
  \bibinfo{author}{\bibfnamefont{P.}~\bibnamefont{Manuel}},
  \bibinfo{author}{\bibfnamefont{S.}~\bibnamefont{Bachus}},
  \bibinfo{author}{\bibfnamefont{F.}~\bibnamefont{Gru\ss{}ler}},
  \bibinfo{author}{\bibfnamefont{P.}~\bibnamefont{Gegenwart}},
  \bibinfo{author}{\bibfnamefont{J.}~\bibnamefont{Singleton}},
  \bibinfo{author}{\bibfnamefont{R.~D.} \bibnamefont{Johnson}},
  \bibinfo{author}{\bibfnamefont{H.~C.} \bibnamefont{Walker}},
  \bibinfo{author}{\bibfnamefont{D.~T.} \bibnamefont{Adroja}},
  \bibinfo{author}{\bibfnamefont{A.~D.} \bibnamefont{Hillier}},
  \bibnamefont{et~al.}, \bibinfo{journal}{Phys. Rev. B}
  \textbf{\bibinfo{volume}{100}}, \bibinfo{pages}{144432}
  (\bibinfo{year}{2019}).

\bibitem[{\citenamefont{Liu et~al.}(2016)\citenamefont{Liu, Yu, and
  Wang}}]{liu_semiclassical_2016}
\bibinfo{author}{\bibfnamefont{C.}~\bibnamefont{Liu}},
  \bibinfo{author}{\bibfnamefont{R.}~\bibnamefont{Yu}}, \bibnamefont{and}
  \bibinfo{author}{\bibfnamefont{X.}~\bibnamefont{Wang}},
  \bibinfo{journal}{Phys. Rev. B} \textbf{\bibinfo{volume}{94}},
  \bibinfo{pages}{174424} (\bibinfo{year}{2016}).

\bibitem[{zha()}]{zhangmeng_unpublished}
\bibinfo{note}{Zheng, Meng and Chao Wang and Yongjian Han and Lixin He,
  unpublished}.

\bibitem[{\citenamefont{Rousochatzakis
  et~al.}(2016)\citenamefont{Rousochatzakis, R\"ossler, van~den Brink, and
  Daghofer}}]{rousochatzakis_kitaev_2016}
\bibinfo{author}{\bibfnamefont{I.}~\bibnamefont{Rousochatzakis}},
  \bibinfo{author}{\bibfnamefont{U.~K.} \bibnamefont{R\"ossler}},
  \bibinfo{author}{\bibfnamefont{J.}~\bibnamefont{van~den Brink}},
  \bibnamefont{and} \bibinfo{author}{\bibfnamefont{M.}~\bibnamefont{Daghofer}},
  \bibinfo{journal}{Phys. Rev. B} \textbf{\bibinfo{volume}{93}},
  \bibinfo{pages}{104417} (\bibinfo{year}{2016}).

\bibitem[{\citenamefont{Becker et~al.}(2015)\citenamefont{Becker, Hermanns,
  Bauer, Garst, and Trebst}}]{becker_spin-orbit_2015}
\bibinfo{author}{\bibfnamefont{M.}~\bibnamefont{Becker}},
  \bibinfo{author}{\bibfnamefont{M.}~\bibnamefont{Hermanns}},
  \bibinfo{author}{\bibfnamefont{B.}~\bibnamefont{Bauer}},
  \bibinfo{author}{\bibfnamefont{M.}~\bibnamefont{Garst}}, \bibnamefont{and}
  \bibinfo{author}{\bibfnamefont{S.}~\bibnamefont{Trebst}},
  \bibinfo{journal}{Phys. Rev. B} \textbf{\bibinfo{volume}{91}},
  \bibinfo{pages}{155135} (\bibinfo{year}{2015}).

\bibitem[{\citenamefont{Zhu et~al.}(2018)\citenamefont{Zhu, Maksimov, White,
  and Chernyshev}}]{zhu_topography_2018}
\bibinfo{author}{\bibfnamefont{Z.}~\bibnamefont{Zhu}},
  \bibinfo{author}{\bibfnamefont{P.~A.} \bibnamefont{Maksimov}},
  \bibinfo{author}{\bibfnamefont{S.~R.} \bibnamefont{White}}, \bibnamefont{and}
  \bibinfo{author}{\bibfnamefont{A.~L.} \bibnamefont{Chernyshev}},
  \bibinfo{journal}{Phys. Rev. Lett.} \textbf{\bibinfo{volume}{120}},
  \bibinfo{pages}{207203} (\bibinfo{year}{2018}).

\bibitem[{\citenamefont{Wu et~al.}(2020)\citenamefont{Wu, Yao, and
  Wu}}]{wu_exact_2020}
\bibinfo{author}{\bibfnamefont{M.}~\bibnamefont{Wu}},
  \bibinfo{author}{\bibfnamefont{D.-X.} \bibnamefont{Yao}}, \bibnamefont{and}
  \bibinfo{author}{\bibfnamefont{H.-Q.} \bibnamefont{Wu}},
  \bibinfo{journal}{arXiv:2008.08751 [cond-mat]}  (\bibinfo{year}{2020}).

\bibitem[{\citenamefont{Verstraete and Cirac}(2004)}]{verstraete04}
\bibinfo{author}{\bibfnamefont{F.}~\bibnamefont{Verstraete}} \bibnamefont{and}
  \bibinfo{author}{\bibfnamefont{J.~I.} \bibnamefont{Cirac}},
  \bibinfo{journal}{cond-mat/0407066}  (\bibinfo{year}{2004}).

\bibitem[{\citenamefont{Verstraete et~al.}(2008)\citenamefont{Verstraete, Murg,
  and Cirac}}]{Verstraete2008}
\bibinfo{author}{\bibfnamefont{F.}~\bibnamefont{Verstraete}},
  \bibinfo{author}{\bibfnamefont{V.}~\bibnamefont{Murg}}, \bibnamefont{and}
  \bibinfo{author}{\bibfnamefont{J.}~\bibnamefont{Cirac}},
  \bibinfo{journal}{Advances in Physics} \textbf{\bibinfo{volume}{57}},
  \bibinfo{pages}{143} (\bibinfo{year}{2008}).

\bibitem[{\citenamefont{Liu et~al.}(2018{\natexlab{b}})\citenamefont{Liu, Dong,
  Wang, Han, An, Guo, and He}}]{Liuwy2018}
\bibinfo{author}{\bibfnamefont{W.-Y.} \bibnamefont{Liu}},
  \bibinfo{author}{\bibfnamefont{S.}~\bibnamefont{Dong}},
  \bibinfo{author}{\bibfnamefont{C.}~\bibnamefont{Wang}},
  \bibinfo{author}{\bibfnamefont{Y.}~\bibnamefont{Han}},
  \bibinfo{author}{\bibfnamefont{H.}~\bibnamefont{An}},
  \bibinfo{author}{\bibfnamefont{G.-C.} \bibnamefont{Guo}}, \bibnamefont{and}
  \bibinfo{author}{\bibfnamefont{L.}~\bibnamefont{He}}, \bibinfo{journal}{Phys.
  Rev. B} \textbf{\bibinfo{volume}{98}}, \bibinfo{pages}{241109(R)}
  (\bibinfo{year}{2018}{\natexlab{b}}).

\bibitem[{\citenamefont{Cao et~al.}(2009)\citenamefont{Cao, Guo, Vanderbilt,
  and He}}]{cao2009first}
\bibinfo{author}{\bibfnamefont{K.}~\bibnamefont{Cao}},
  \bibinfo{author}{\bibfnamefont{G.-C.} \bibnamefont{Guo}},
  \bibinfo{author}{\bibfnamefont{D.}~\bibnamefont{Vanderbilt}},
  \bibnamefont{and} \bibinfo{author}{\bibfnamefont{L.}~\bibnamefont{He}},
  \bibinfo{journal}{Phys. Rev. Lett.} \textbf{\bibinfo{volume}{103}},
  \bibinfo{pages}{257201} (\bibinfo{year}{2009}).

\bibitem[{\citenamefont{Kawamura and Miyashita}(1984)}]{kawamura84}
\bibinfo{author}{\bibfnamefont{H.}~\bibnamefont{Kawamura}} \bibnamefont{and}
  \bibinfo{author}{\bibfnamefont{S.}~\bibnamefont{Miyashita}},
  \bibinfo{journal}{J. Phys. Soc. Jpn.} \textbf{\bibinfo{volume}{53}},
  \bibinfo{pages}{4138} (\bibinfo{year}{1984}).

\bibitem[{\citenamefont{Li et~al.}(2020)\citenamefont{Li, Liao, Chen, Zeng,
  Sheng, Qi, Meng, and Li}}]{Lihan2020}
\bibinfo{author}{\bibfnamefont{H.}~\bibnamefont{Li}},
  \bibinfo{author}{\bibfnamefont{Y.~D.} \bibnamefont{Liao}},
  \bibinfo{author}{\bibfnamefont{B.-B.} \bibnamefont{Chen}},
  \bibinfo{author}{\bibfnamefont{X.-T.} \bibnamefont{Zeng}},
  \bibinfo{author}{\bibfnamefont{X.-L.} \bibnamefont{Sheng}},
  \bibinfo{author}{\bibfnamefont{Y.}~\bibnamefont{Qi}},
  \bibinfo{author}{\bibfnamefont{Z.~Y.} \bibnamefont{Meng}}, \bibnamefont{and}
  \bibinfo{author}{\bibfnamefont{W.}~\bibnamefont{Li}}, \bibinfo{journal}{Nat.
  Commun.} \textbf{\bibinfo{volume}{11}}, \bibinfo{pages}{1111}
  (\bibinfo{year}{2020}).

\end{thebibliography}

\end{document}